**Realization of electron vortices with large orbital angular momentum using miniature holograms fabricated by electron beam lithography**

E. Mafakheri[1], A. H. Tavabi[2], P.-H. Lu[2], R. Balboni[3], F. Venturi[1,4], C. Menozzi[1,4], G. C. Gazzadi[4], S. Frabboni[1,4], A. Sit[5], R. E. Dunin-Borkowski[2], E. Karimi[5,6], V. Grillo[4,7,*]

1. Dipartimento di Fisica Informatica e Matematica, Università di Modena e Reggio Emilia, via G Campi 213/a, I-41125 Modena, Italy.
2. Ernst Ruska-Centre for Microscopy and Spectroscopy with Electrons (ER-C) and Peter Grünberg Institute (PGI), Forschungszentrum Jülich, D-52425 Jülich, Germany.
3. CNR-IMM, Via P. Gobetti 101, I-40129 Bologna, Italy.
4. CNR-Istituto Nanoscienze, Centro S3, Via G Campi 213/a, I-41125 Modena, Italy.
5. Department of Physics, University of Ottawa, 25 Templeton, Ottawa, Ontario, K1N 6N5 Canada.
6. Department of Physics, Institute for Advanced Studies in Basic Sciences, 45137-66731 Zanjan, Iran.
7. CNR-IMEM Parco Area delle Scienze 37/A, I-43124 Parma, Italy.
*  Corresponding author: vincenzo.grillo@unimore.it

**Abstract**:

Free electron beams that carry high values of orbital angular momentum (OAM) possess large magnetic moments along the propagation direction. This makes them an ideal probe for measuring the electronic and magnetic properties of materials, and for fundamental experiments in magnetism. However, their generation requires the use of complex diffractive elements, which usually take the form of nano-fabricated holograms. Here, we show how the limitations of focused ion beam milling in the fabrication of such holograms can be overcome by using electron beam lithography. We demonstrate experimentally the realization of an electron vortex beam with the largest OAM value that has yet been reported (L = 1000ℏ), paving the way for even more demanding demonstrations and applications of electron beam shaping.

Similarly to its optical counterpart [1], an electron vortex beam (EVB) possesses one or more phase singularities at the center of its helical wavefront, and is an eigenstate of the component of orbital angular momentum (OAM) along its propagation direction with eigenvalue $\ell\hbar$ (where $\ell$ is an integer and $\hbar$ is the reduced Planck constant) [2-5]. As an electron is a charged particle, an EVB has a magnetic moment of $\ell\mu_B$, where $\mu_B$ is the Bohr magneton. Both its magnetic moment and its angular momentum allow for coupling to materials and intriguing applications, including magnetic and shape dichroism measurements [6-9], chiral crystal structure characterization[10], nanoparticle manipulation[11] and electron spin polarization[12]. EVBs are also of fundamental interest as they are characterized by a discrete quantum number that can form the basis of quantum experiments[13]. For values of $\ell$ of a few units, the resulting magnetic effects are of the same order of magnitude as spin effects. However, the magnetic moment increases linearly with $\ell$ and can in principle be orders of magnitude larger, since there is no fundamental upper bound for $\ell$.

The realization of a high OAM value is of great importance for the amplification of subtle physical effects. For example, a magnetic component of transition radiation has been predicted for large OAM beams[14]. They have also been proposed for the measurement of out-of-plane magnetic fields in nanostructures using transmission electron microscopy (TEM) through the Larmor/Zeeman interaction[15,16]. Moreover, large electron vortex beams are interesting quantum objects in their own right. Whereas the TEM electron wavelength is typically on the order of 2 pm, a single highly twisted wavefront winds up with a step length of up to a few nanometers (Figure 1a). By following a single wavefront as it spirals along its propagation direction, one would therefore find the same azimuth after a few nanometers. The realization of such a characteristic length in the longitudinal direction on the nanometer scale (*i.e.*, a scale that is typical for nanomaterial experiments) is highly desirable.

Finally, electron vortex beams can be coupled to Landau states in the magnetic lens of a TEM (a longitudinal magnetic field). Landau states possess a functional similarity to the class of EVBs that are termed Laguerre-Gaussian beams and are characterized by a spiraling phase corresponding to $L = \ell\hbar$ and a radial index $p$ [13,17,18]. The transverse energies of such states can be written in the form

$$\varepsilon = \hbar\Omega(2p + \ell + |\ell| + 1), \qquad (1)$$

where $\Omega = eB/2m$ is the Larmor frequency, B is the magnetic field and *m* and *e* are the electron mass and charge, respectively. Neglecting for the moment the large spread over the *p* degree of freedom of most electron beams (*e.g.*, in Ref. [19] we have shown an extreme case of dispersion in *p* decomposition for an EVB), for a typical magnetic field B of 2 T inside an electron microscope, the discrete transverse energy of an excited state corresponding to a few 1000 ℏ can be as high as 0.5 to 1 eV, and can therefore potentially be coupled to infrared/visible light (Figure 1b).

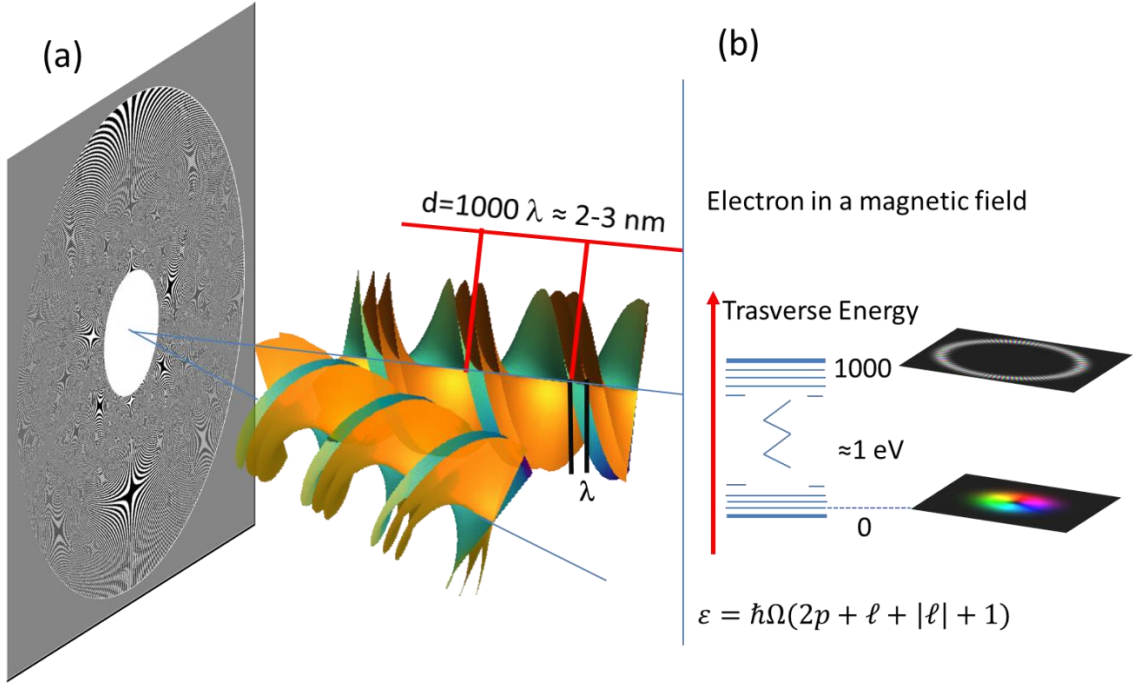

**Figure 1** (a) Schematic diagram of the generation of high order vortex beams. A hologram diffracts an electron beam, producing two primary opposite vortices. For typical wavelengths in the TEM of between 2 and 3 pm and a vortex with ℓ =1000, the step-length (*i.e.*, the distance between two successive arrivals of the wavefront to the same azimuth) is on the order of a few nm. (b) Schematic diagram showing the transverse energy for Landau states. Laguerre-Gauss modes would couple to Landau states in the presence of a magnetic field, giving rise to discrete transverse energy states. For ℓ = 1000, the energy difference to the ground state is on the order of some 0.1 eV.

Unfortunately, the experimental realization of large OAM electron vortices has been hindered technically by the approaches to nano-fabrication that have been used. Since EVBs were predicted theoretically[2], several different methods have been used to generate them, involving the use of spiral phase plates[3,20-22], pitch-fork holograms[4,5,19,23,24], spiral zone plates[25,26], Hilbert phase plates coupled to quadrupole lenses[27], multipole lenses in aberration correctors[28] and both magnetic[28,29] and electrostatic[30] phase plates. From these options, off-axis phase holograms[5,19,23,24,32] are still the method of choice. In such holograms, phase changes are introduced in proportion to their local thickness. EVBs have been reported with values for $\ell$ of 100-200$\hbar$ in high diffraction orders[5] and a first-order value for $\ell$ of 200$\hbar$ with higher efficiency[19]. Such phase holograms can be fabricated with different groove profiles to produce different intensity distributions in their diffraction orders[23]. For example, a blazed triangular profile can potentially be used to convey all of the transmitted intensity into a single diffracted beam[23,33]. Other less challenging and more popular groove profiles take sinusoidal and rectangular forms. For a perfectly tuned thickness, a rectangular shape is superior to a sinusoidal shape, as it inhibits all even diffraction orders, including the $0^{th}$ order.

The limited resolution of focused ion beam (FIB) milling, which has been used to fabricate the holograms in these examples, prevents the realization of higher OAM beams since the primary grating periodicity must be decreased to achieve an increasing OAM in order to avoid the superposition of electron vortices of different diffraction orders. A limited resolution can also mean that an intended rectangular groove profile can end up being nearly sinusoidal. Moreover, it is demanding to maintain the same groove profile and a uniform response over both high and low spatial frequencies. Additional problems include the total patterning time and the total number of addressable pixels.

Here, we overcome these limitations by using different protocols based on electron beam lithography (EBL) to allow us to achieve a vortex with a topological charge as high as 1000 $\hbar$. EBL is a technique that is used widely to produce patterns based on the selective electron irradiation of an electron sensitive material. We use a Zeiss Σ scanning electron microscope equipped with a Schottky field emitter and a Raith Elphy Quantum pattern generator.

In order to optimize the spatial resolution of EBL patterning, we tested both positive and negative resists. Even though EBL is widely used for device fabrication, it is less frequently applied on thin SiN membranes in the form of 3 mm disk-shaped TEM specimens[34]. We used square 50-nm-thick SiN membranes, on which the electron-transparent region had a width of 80 μm. The membranes were covered with evaporated Au (typically 200 nm thick) that was removed only in the hologram region. The procedure required 2 steps of lithography. A first step was used to create an electron transparent aperture in the Au mask in the active region of the hologram. A second step involved patterning the holograms with appropriate phase modulations. The pattern thickness was defined according to the formula

$$t = \frac{1}{2}t_0(1 + sign(\sin(\ell\theta + \rho k_{carrier} \cos(\theta)))) , \qquad (2)$$

where $\rho, \theta$ are polar co-ordinates in the hologram plane, $k_{carrier}$ is the carrier frequency in the off-axis hologram, and *sign[.]* is the sign function, which is ±1 for positive and negative arguments, respectively. The thickness $t_0$ was chosen to provide a phase difference close to π. The rectangular groove shape defined by eq 2 conveniently allowed the use of EBL to produce holograms with 2 discrete thickness levels.

To first order, the separation *d* between two hologram lines is related to the argument of the sin function, *i.e.*, $f = \ell\theta + \rho k_{carrier} \cos(\theta)$, through the expression to the first order $\frac{1}{d} \approx \frac{|\nabla_{\rho,\theta}(f)|}{\pi} + \cdots$. When the function *f* is stationary (*i.e.*, when $\nabla_{\rho,\theta}(f) = 0$), the separation between the lines increases. This condition is realized when $\rho = \ell/k_{carrier}, \theta = \frac{\pi}{2}$. A detailed analysis shows that the *f*-stationary point is a saddle point for *f*. Such a saddle point is visible in Figure 1 and in our previous work for ℓ = 200, taking the form of a cross close to the center of the hologram[19]. The lines in the hologram have a much higher frequency in the center and opposite the f-stationary point. As in the previous paper, we decided to exclude the central region of the hologram from patterning[5,19]. For all patterning, a bitmap image was created using the STEMCELL software package[35] and converted to a data format that was readable using the EBL pattern generator.

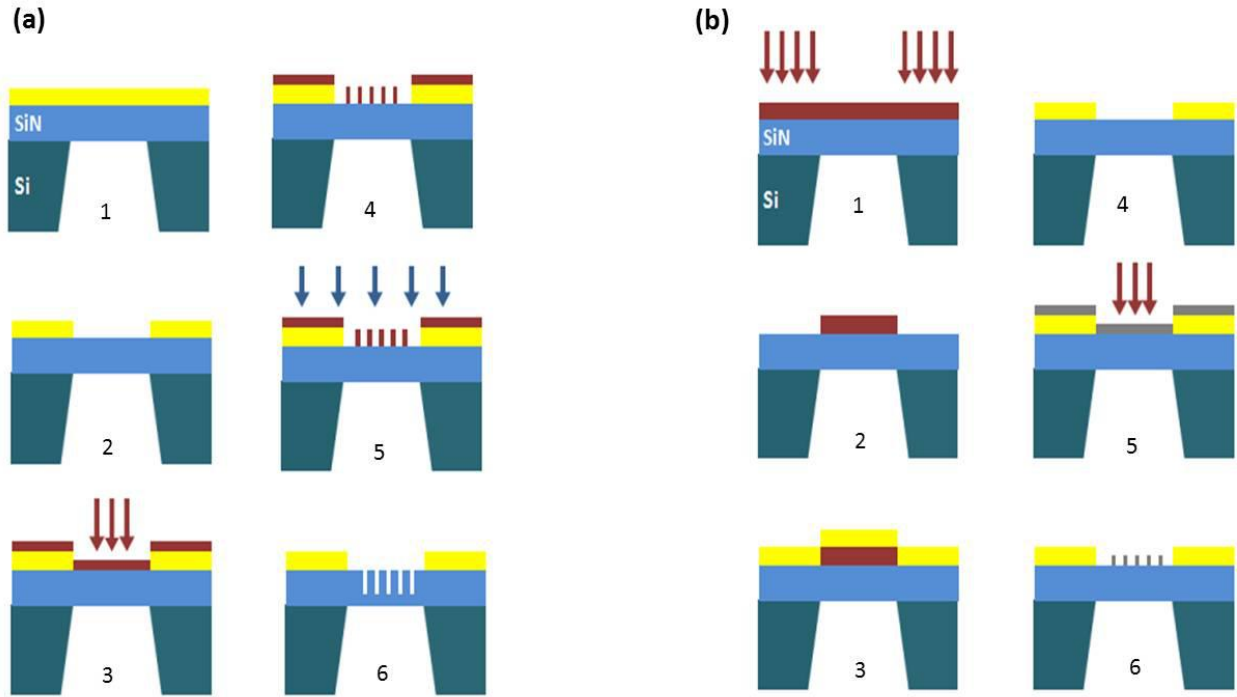

**Figure 2** (a) Schematic diagrams showing (1) the evaporation of a Au layer; (2) the creation of a Au aperture using FIB milling; (3) PMMA spin coating and lithography; (4) developing the pattern; (5) reactive ion etching; (6) removing the resist. (b) Schematic diagrams showing (1) spin coating of PMMA and EBL; (2) developing the resist; (3) Au evaporation; (4) lift-off; (5) hydrogen silsesquioxane (HSQ) spin coating and EBL; (6) developing the HSQ.

Figure 2a shows the lithography method that we first used for the positive resist. In this case, we used polymethyl methacrylate (PMMA) for patterning the hologram, and reactive ion etching (RIE) to transfer the pattern onto the SiN membrane. This process suffered from limitations in spatial resolution, primarily due to the resolution that could be achieved using PMMA in our instrument. As a result, it did not provide a real improvement over FIB patterning. In contrast, Figure 2b shows the process that gave the best results in terms of resolution. In this case, we used hydrogen silsesquioxane (HSQ), which transformed into a silica-like structure after baking on a hot plate. The baked HSQ, which was resistant to electron beam irradiation, was the material that was used to impart a phase difference to the electron wave in the TEM. The use of such a negative resist provided considerable advantages in terms of ease of use, both because no RIE step was necessary (requiring only development of the written pattern) and because of the superior resolution of the final pattern. The only disadvantage was the insulating nature of the HSQ, which introduced a large charging effect during TEM examination. For this reason, each hologram was coated with a few nanometers of evaporated Cr. This approach solved most of the charging problems. The result of this patterning is shown in the form of a TEM image of a hologram (for a version without Au) in Figure 3a. The patterned area is clearly visible, as is the large central hole. In the final version, we did not cover this central part with Au[19] but left it unpatterned, with Au still present in the external regions. On the left, a region of stationary phase is visible (indicated by a circle). Although the image appears to show other stationary points, these are artifacts resulting from digital reproduction due to undersampling of the TEM image. Figure 3b shows a TEM image of part of the hologram, which confirms that different spatial frequencies are reproduced correctly. Figure 3c shows a thickness map, calculated using energy-filtered TEM [19], of part of the pattern, in which line widths of 18 nm are present and the average periodicity is below 65 nm, *i.e.*, approximately half of the best typical FIB resolution achieved when fabricating similar holograms. The map indicates that, even at such a scale, the lateral definition of the trench is good, that the vertical step is nearly perfect and that the trench thickness is uniform.

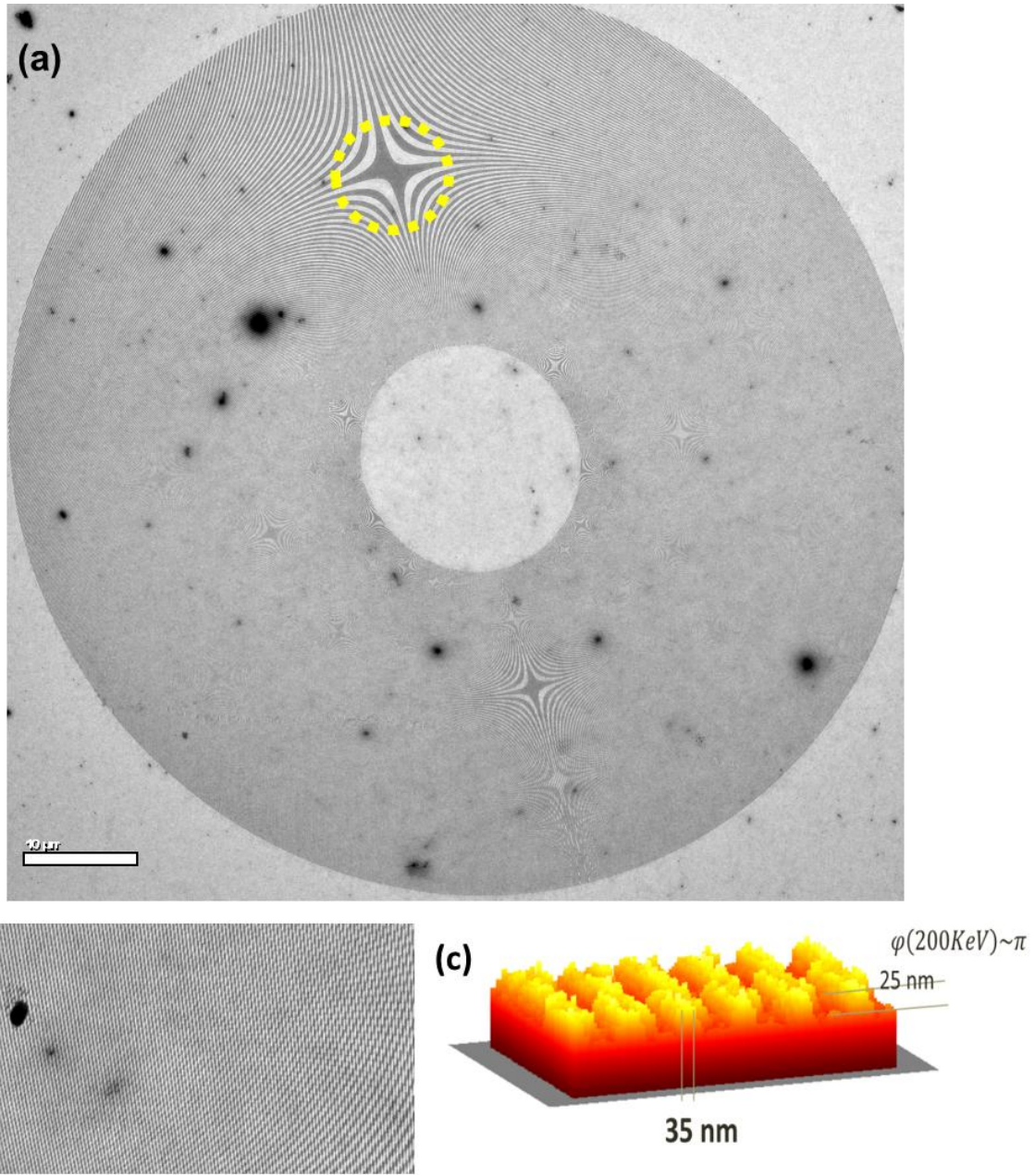

**Figure 3** (a) TEM image of an HSQ-based hologram for ℓ = 1000ℏ. (b) Higher magnification TEM image of a region of the hologram, showing both low and high spatial frequencies in the pattern. (c) 3D rendering of an energy-filtered-TEM-based thickness map of a region of one of the holograms, showing detail on the order of 35 nm.

Figure 4a shows a nearly-in-focus image of a Fraunhofer diffraction pattern of the hologram recorded at 300 keV using a FEI Titan equipped with a Schottky FEG and operated in LowMag mode. The two opposite vortices take the form of rings with uniformly bright intensities, confirming the very small proximity effect in our EBL pattern, *i.e.*, that both high and low spatial frequencies are transferred correctly to the electron wave. We also observe only a very faint trace of second order diffraction, providing clear confirmation that the grooves are sharp and almost rectangular in shape (as suggested by the thickness map) and that the phase difference between the thin and thick region almost perfectly matches the intended value of π.

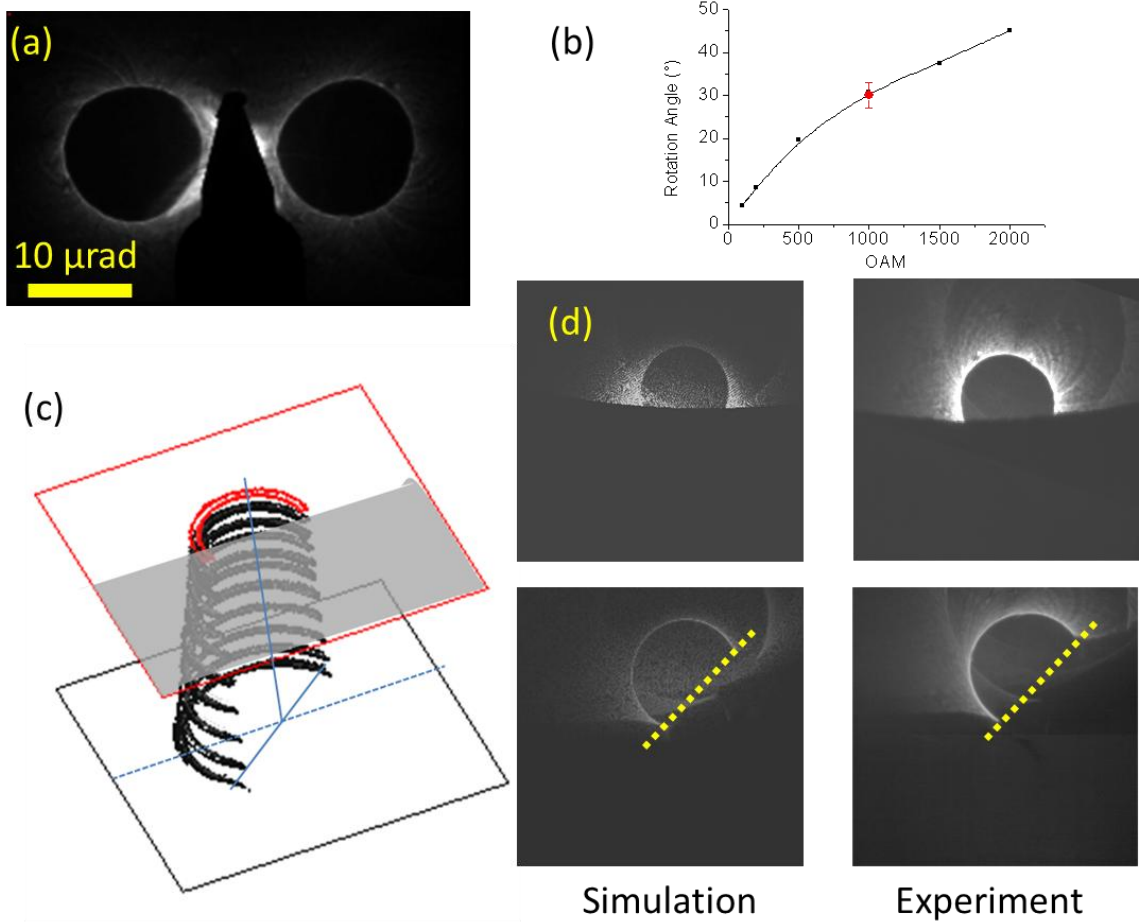

**Figure 4** (a) Experimental image of a full diffraction pattern of the hologram nearly in focus. A beam stop was used to block the transmitted beam. (b) Plot of the expected rotation as a function of orbital angular momentum quantum number, shown together with our experimental measurement. (c) Schematic diagram illustrating the use of a knife edge to measure beam rotation. (d) Experimental results and simulations for ℓ=1000 before and after propagation.

We now address measurement of the OAM value of the vortex. In a previous article[29], we simply measured the thickness/phase in the hologram plane. Here, considering the large size of the pattern with respect to those fabricated previously[23], we are not able to make a reasonable thickness map of the entire hologram (we would need an digital image with a size of at least a $4\times10^8$ pixels). Therefore, we cannot check the exact OAM spectrum of the vortex in this way.

Instead, inspired by a suggestion in the literature[35-37], we used the selected area diffraction (SAD) aperture to block half of the vortex. We then systematically varied the excitation of the diffraction lens, which is located immediately after the aperture. In this way, we observed the rotation of the Fresnel diffraction image. A simplified way to calculate this rotation angle makes use of the expression

$$\theta = (\frac{eB}{2m} \pm \frac{L}{mr^2})\frac{\Delta z}{v}, \qquad (3)$$

where $\Delta z$ is the propagation distance, $v$ is the electron velocity and the quantity $r$ is a "semi-classical" value of the radius, whose correct quantum interpretation depends on the shape of the beam[37]. The expression for the rotation angle contains two terms. The first is the Larmor rotation, which depends on the magnetic field of the diffraction lens. The second is referred to as the Gouy rotation and is associated with the phase gradient of the vortex. As the Gouy rotation goes to zero at large distances $r$ from the rotation center, we can separate the Larmor contribution by checking the rotation of the shadow of the aperture far from the center. For our large OAM vortex, as a result of the use of a weak diffraction lens, this contribution is small (10% of the overall rotation), while the dominant contribution to the rotation originates from the phase gradient of the vortex itself.

For an exact evaluation of the Gouy rotation, instead of using eq 3, we evaluated the Fresnel propagation numerically for different values of ℓ. We calibrated the defoci by comparing simulations and experimental images recorded without an aperture, since the absolute $z$ positions of both the SAD aperture and the Fraunhofer diffraction pattern were unknown. We define the parameter

$$z_R = \pi \frac{r_{rim}^2}{\lambda \cdot 1000}, \tag{4}$$

which would be the Rayleigh range for an ideal Laguerre-Gauss beam with $p = 0$ and an equivalent apparent rim radius $r_{rim}$ (the radius corresponding to maximal intensity). (See the Supplementary Material). We find that the aperture is located at $z/z_R = 0.5$ and that we analyzed the rotation after $z/z_R = 2$. Figure 4b shows the expected rotation according to simulations for beams of different values of L and for the realistic hologram structure. The best match for <L> is (960±120) $\hbar$ and is consistent with the nominal value. Figure 4c shows a comparison between a simulation for $L = 1000\hbar$ and our experimental results, demonstrating good agreement.

To conclude, we have demonstrated that by using EBL we can overcome intrinsic limitations in the creation of nano-fabricated holograms when compared to using FIB milling, in terms of 1) the maximum OAM that can be reached; 2) the minimum detail that can be reproduced (reaching a spatial resolution of at least 33 nm); 3) improved uniformity of the frequency response; 4) better suppression of higher order diffraction due to a nearly perfect rectangular groove profile. We believe that EBL will be the fabrication technique of choice for future complex diffractive optics with electrons. By using a very large number of pixels and a very small pitch, a large separation between the diffracting orders and good definition of the phase modulation can be achieved. One of the most interesting perspectives is to use holograms to shape beams that have very well defined and stable properties, in order to increase the precision of TEM measurements. For example, the case of $L = 1000\hbar$ can be used to increase the sensitivity of vertical magnetic field measurements. A very extended and precise grating can be used for electron interferometry, while by using an extended version of a conical hologram[32], it will be possible to create a nearly ideal Bessel beam or a perfectly narrow ring shaped beam, both of which are potentially useful for interferometry or for measurements of small deflections. The OAM of $L = 1000\hbar$ that we have demonstrated is the largest value that has been achieved so far. Beyond magnetic measurements, we will use it to verify coupling with Landau states and for the observation of the, until now elusive, OAM-dependent transition radiation[14].

**Acknowledgements**

V.G. acknowledges the support of the Alexander von Humboldt Foundation. We thank P. Pingue and F. Carillo for the access to the lithography facility of CNR.Nest in Pisa (Italy). S.F. and F.V.

thanks the University of Modena and Reggio Emilia for the grant FAR-2015-Project Title: Computer Generated Holograms for the realization and analysis of structured electron waves. The research leading to these results has received funding from the European Research Council under the European Union's Seventh Framework Programme (FP7/2007-2013)/ ERC grant agreement number 320832. E.K. and A.S. acknowledges the support of the Canada Research Chairs (CRC) and Canada Foundation for Innovations (CFI) Programs.